%% file: paper.tex
\documentclass[11pt]{article}
\usepackage{graphicx}

\newcommand{\BABARPubYear}    {04}

\newcommand{\BABARConfNumber} {42}
\newcommand{\SLACPubNumber} {10634}

\def\BtoKpipi{\Bz\to K^+\pi^-\pi^0}

\def\dE{{\Delta E}}
\def\de{\Delta E}
\def\deprime{{\de^\prime}{}}
\def\mes{\ensuremath{m_{\rm ES}}}

\def\Amptp{{\cal A}_{B^0 \rightarrow K^+\pi^-\pi^0}}
\def\Amptpbar{\kern 0.18em\overline{\kern -0.18em {\cal A}}_{{\overline B^0} \rightarrow K^-\pi^+\pi^0}}
\def\Anonres{{\cal A}_{\rm NR}}
\def\Ares{{\cal A}}
\def\fscfave{\kern 0.18em\overline{\kern -0.18em f}_{\rm SCF}}

\def\TM{{\rm TM}}
\def\SCF{{\rm SCF}}

\def\QKi{q_{{\rm K},i}}
\def\ie{{\em i.e.}}

\def\AmpAll{|\Amptp|^2+|\Amptpbar|^2}
\def\fscfi{f_{{\rm SCF},i}}
\def\ea{{\em et al.}}
\def\Ares{{\cal A}}
\def\DP{{\rm DP}}%

\newcommand{\beq}{\begin{equation}}
\newcommand{\eeq}{\end{equation}}
\newcommand{\beqn}{\begin{eqnarray}}
\newcommand{\eeqn}{\end{eqnarray}}

\input pubboard/babarsym

\long\def\inst#1{\par\nobreak\kern 4pt\nobreak
    {\it #1}\par\vskip 10pt plus 3pt minus 3pt}

\begin{document}
{\pagestyle{empty}

\begin{flushright}
\babar-CONF-\BABARPubYear/\BABARConfNumber \\
SLAC-PUB-\SLACPubNumber \\
August 2004 \\
\end{flushright}

\par\vskip 5cm

\begin{center}
\Large \bf $B^0 \rightarrow K^+ \pi^- \pi^0$ Dalitz Plot Analysis
\end{center}
\bigskip

\begin{center}
\large The \babar\ Collaboration\\
\mbox{ }\\
\today
\end{center}
\bigskip \bigskip

\begin{center}
\large \bf Abstract
\end{center}
We present preliminary results on the Dalitz plot analysis of $B^0
\rightarrow K^+ \pi^- \pi^0$ decays. The data sample comprises 213 
million $\FourS \to B\Bbar$ decays collected with the \babar\ detector 
at the \pep2 asymmetric-energy $B$~Factory at SLAC. We report 
measurements of the inclusive branching fraction, quasi-two-body 
fractions and \CP-violating 
charge asymmetries for intermediate states including 
$K^*(892)^+ \pi^-$ and $\rho(770)^- K^+$. Observations of $B^0$ decays 
to the $K\pi$ $S$-wave intermediate states, $K_0^*(1430)^+ \pi^-$ and 
$K_0^*(1430)^0 \pi^0$, are reported. 
Evidence of the decay $B^0 \rightarrow K^*(892)^0 \pi^0$ is seen.
We set upper limits at 90\% confidence level on branching fractions of 
the nonresonant and other less significant intermediate states.

\vfill
\begin{center}

Submitted to the 32$^{\rm nd}$ International Conference on High-Energy Physics, ICHEP 04,\\
16 August---22 August 2004, Beijing, China

\end{center}

\vspace{1.0cm}
\begin{center}
{\em Stanford Linear Accelerator Center, Stanford University, 
Stanford, CA 94309} \\ \vspace{0.1cm}\hrule\vspace{0.1cm}
Work supported in part by Department of Energy contract DE-AC03-76SF00515.
\end{center}

\newpage
} 

\input pubboard/authors_sum2004.tex

\section{INTRODUCTION}
\label{sec:Introduction}
  Charmless three-body $B$ decays significantly broaden the 
understanding of
$B$ meson decay mechanisms and provide additional possibilities for direct
\CP\ violation searches. The $B^0 \rightarrow K^+ \pi^- \pi^0$  
decay is known to have contributions from the 
charmless intermediate $B^0 \rightarrow K^*(892)^+\pi^-$ 
and $B^0 \to\rho(770)^-K^+$ decays. 
Although not yet observed, $B^0\rightarrow K^*(892)^0\pi^0$ could 
also contribute, as well as $K_0^*(1430)$, 
$K_2^*(1430)$, $K^*(1680)$, $\rho(1450)$, and $\rho(1700)$.
The $B^0 \rightarrow K^+ \pi^- \pi^0$ decay can also occur via long-lived 
charmed intermediate states, \ie\,, $B^0 \rightarrow {\overline D^0} \pi^0$,
and the doubly-Cabibbo-suppressed $B^0 \rightarrow D^- K^+$,
$D^- \rightarrow \pi^- \pi^0$. 
Intermediate states with broad resonances are interfering and therefore
a full amplitude analysis is required to extract the amplitudes and 
relative phases. Quasi-two-body 
branching fractions and $CP$-violating charge asymmetries for 
intermediate states are also measured. The results can be used to
probe the weak phase 
$\gamma \equiv \arg\left[-V_{ud}^{}V_{ub}^{*}/V_{cd}^{}V_{cb}^{*}\right]$
of the Unitarity Triangle~\cite{CKM, Gronau}.

  The charmless intermediate states are largely dominated by a gluonic
penguin diagram ($\bar{b} \rightarrow u \bar{u} \bar{s}$ or 
$d \bar{d} \bar{s}$),
while the electroweak penguin contribution may also be 
sizeable~\cite{Deshpande_PEW}. The Dalitz plot analysis provides a 
sensitive probe on the penguin contributions to $B$ decays to a 
charmless vector meson and a charmless pseudoscalar meson.


Measurements of some of the branching fractions and charge asymmetries have
been carried out by CLEO~\cite{CLEOkpipi0}, 
\babar~\cite{BaBarRhopi} and Belle~\cite{BELLEkpipi0}
in the quasi-two-body approximation.
In this report, we present a Dalitz plot analysis that takes into account
interference between individual modes,
using a Maximum Likelihood (ML) approach. Unless stated otherwise, 
charge conjugation is always implied throughout this paper.

\section{THE \babar\ DETECTOR AND DATASET}
\label{sec:babar}
The data used in this analysis were collected with the \babar\ detector
at the \pep2\ asymmetric-energy $e^+e^-$ storage ring. 
The sample consists of 213 million
$B{\overline B}$ pairs, corresponding to an integrated luminosity of
193.2~$\invfb$ collected at the $\Upsilon(4S)$ resonance (``on-resonance''),
 and an integrated luminosity of 16~\invfb collected about
40~\mev below the \FourS (``off-resonance'').

The \babar\ detector is described in detail elsewhere~\cite{ref:babar}.
A five-layer double-sided silicon vertex tracker (SVT) and a 40-layer 
drift chamber (DCH) are used to detect charged particles and measure 
their momentum as well as ionization energy loss ($dE/dx$). 
Charged hadrons are identified with a detector of internally reflected
Cherenkov light (DIRC) and  ionization in the tracking detectors. 
Photons, neutral hadrons, and electrons are detected in a CsI(Tl) 
calorimeter (EMC), while muons are identified
in the magnetic flux return system (IFR).

\section{ANALYSIS METHOD}
\label{sec:Analysis}
\subsection{Event Selection}
Each signal $B$ candidate is reconstructed from $K^+$, $\pi^-$, and $\pi^0$ 
candidates. Charged tracks must have at least 12 hits in the DCH 
and a transverse momentum larger than $100~\mevc$. 
Charged tracks identified as electrons, muons or protons are rejected.
We also require that the charged kaon candidate must be identified 
as a kaon and that the charged pion candidate must not be identified
as a kaon.  
The $\pi^0$ candidate must have a mass that satisfies
$0.11<m(\gamma\gamma)< 0.16 \gevcc$, and each photon is required to have 
an energy greater than $50~\mev$ in the laboratory frame and to exhibit a 
lateral profile of energy deposition in the EMC consistent with an 
electromagnetic shower.

Two kinematic variables, $\dE$ and $\mes$, allow the discrimination 
of signal \B decays from random combinations of tracks and
$\pi^0$ candidates. The energy difference,
$\dE$, is the difference between the $\epem$ center-of-mass (CM)
energy of the \B candidate and $\sqrt{s}/2$, where $\sqrt{s}$ is the 
total CM energy. The beam-energy-substituted mass, $\mes$, is defined by
$\sqrt{(s/2+{\mathbf {p}}_i\cdot{\mathbf{p}}_B)^2/E_i^2-{\mathbf {p}}_B^2},$
where the $B$ momentum, ${\mathbf {p}}_B$, and the four-momentum of the 
initial $\epem$ state 
($E_i$, ${\mathbf {p}}_i$) are measured in the laboratory frame.
For the signal $B$ candidate, we require that $\mes > 5.27 \gevcc$.
The $\dE$ resolution depends on the \piz energy and therefore
varies across the Dalitz plot.
To account for this effect, we introduce a transformed quantity:
\begin{equation}
\deprime \equiv \frac{2\dE - (\dE_{max}+\dE_{min})}{\dE_{max} - \dE_{min}}\,,
\end{equation}
where $\dE_{max} = 0.08 - 0.0014 \cdot m(K^+\pi^-)$,
$\dE_{min} = -0.14 + 0.0038 \cdot m(K^+\pi^-)$, all
in units of $\gev$. We require $|\deprime| \leq 1$.

Continuum $e^+e^-\to q\bar{q}$ ($q = u,d,s,c$) events are the dominant 
background. To enhance discrimination between signal and continuum, 
we utilize a neural network (NN) which combines four discriminating 
variables: the cosine of the angle between the \B momentum and the $e^+$
beam direction in the CM frame, the cosine of the angle between 
the \B thrust axis and the beam direction in the CM frame, and
two event-shape variables 
defined as sums over all particles $i$ of $p_i\times|\cos{\theta_i}|^n$, 
where $n=0$ or $2$,
and $\theta_i$ is the angle between the momentum of particle 
$i$ and the $B$ thrust axis.
The NN weights the discriminating variables differently, according to 
training on  off-resonance data and
simulated signal events. We cut on the NN output to suppress 
about 95\% of the continuum background while retaining 62\% of the signal.

The fraction of events that have more than one candidate satisfying the selection
depends on the kinematical distributions of the \B decay products, and is therefore
different for each mode. When more than one $\pi^0$ candidate is present in the event, 
we choose the candidate with the reconstructed $\pi^0$ mass
closest to the nominal value of 0.1349~$\gevcc$. If more than one charged track combination
satisfies the selection, we randomly choose one \B candidate. 
An event is classified as misreconstructed signal if it 
contains a \B 
which decays to the signal mode, but one or more reconstructed particles
are not actually from the decay of that \B. Misreconstructed signal is called 
self cross feed (SCF).
This misreconstruction is primarily due to the presence of low momentum 
pions in the $B$ decays and thus the SCF fraction varies 
across the Dalitz plot.

There are 7220 events selected from the data sample. 
Table~\ref{tab:finalEff} summarizes the signal efficiencies
(including the misreconstructed events) and the misreconstruction fractions
(with respect to the total signal) estimated from simulation with 
non-interfering amplitudes.

\begin{table}[hbt]
\begin{center}
\caption[Final signal selection efficiencies]
        {\label{tab:finalEff}
        Signal efficiencies (misreconstructed events (SCF) included),
        fractions of misreconstructed events, for the different intermediate
        signal modes. The errors given are statistical only.}
\setlength{\tabcolsep}{1.2pc}
\begin{tabular}{lcc} \hline\hline
 Decay mode & $\varepsilon(\%)$ 
& $f_{\rm SCF}(\%)$ \\\hline
&& \\ [-0.2cm]
$K^*(892)^+\pi^-$ & 14.9 $\pm$ 0.1  & 23.6 $\pm$ 0.3 \\
$\rho^- K^+$      & 16.7 $\pm$ 0.0  & 22.2 $\pm$ 0.1 \\
$K^*(892)^0\pi^0$ & 16.8 $\pm$ 0.1  
                  & \phantom{1}8.7 $\pm$ 0.2 \\
$nonresonant$ 
                  & 16.3 $\pm$ 0.0  
                  & \phantom{1}3.8 $\pm$ 0.0 \\
$K_0^*(1430)^+\pi^-$ & 15.5 $\pm$ 0.1  
                  & \phantom{1}9.0 $\pm$ 0.2 \\
$K_0^*(1430)^0\pi^0$ & 17.4 $\pm$ 0.1  
                  & \phantom{1}4.2 $\pm$ 0.2
\\[0.1cm]
\hline\hline
\end{tabular}

\end{center}
\end{table}


We use  simulated events to study the background
from other $B$~decays (\B-background) which includes both 
charmed ($b \to c$) and charmless decays.
In the selected $B^0 \rightarrow K^+ \pi^-\pi^0$ sample we expect
$503\pm151$ $b \to c$ background events.
The decays $B^0 \rightarrow {\overline D^0} \pi^0, 
{\overline D^0} \rightarrow K^+ \pi^-$ and 
$B^0 \rightarrow D^- K^+, D^- \rightarrow \pi^- \pi^0$ 
have the same final state as the charmless intermediate states
of our interest, but they do not interfer with the charmless 
intermediate states. Both are technically treated as 
$B$-backgrounds, but separately from other 
charmed background events.
Branching fractions of $B^0 \rightarrow {\overline D^0} \pi^0$ 
and $B^0 \rightarrow D^- K^+$ have been already measured,
albeit with large uncertainties. In the present analysis, while we still 
do not have enough sensitivity to observe the doubly-Cabibbo suppressed 
$B^0 \rightarrow D^- K^+$, $D^- \to \pi^- \pi^0$, we measure the 
branching fraction of the $B^0 \rightarrow {\overline D^0} \pi^0$ 
decay.
The selection efficiency for this mode is
($16.7 \pm 0.1$)\% with a misreconstruction fraction of $(2.6 \pm 0.1)\%$.
We expect $230\pm20$ charmless \B-background events in the data sample. 
The major charmless \B-background modes include 
$B^0\rightarrow K^*(892)^0\gamma$, $K^*(1430)^0\gamma$, 
$B^0\rightarrow K^+\pi^-$, $B^+\rightarrow K^+\pi^0$, 
$B^0\rightarrow \rho^+\pi^-$,
$B^0\rightarrow \rho^+\rho^-$, and $B^+\rightarrow \rho^+\rho^0$.
Other contributing modes include
$B^+\rightarrow K^*(892)^+\pi^0$, $B^+\rightarrow f^0 K^+$,
$B^+\rightarrow K_0^{*}(1430)^{0}\pi^+$,
and $B^+\rightarrow \eta' K^+$.
The branching fractions of unmeasured decay channels are estimated 
within conservative error ranges.
Charmless \B-background modes are grouped into six classes
with similar kinematic and topological properties.
Two additional classes account for the inclusive neutral and charged 
$b \to c$ decays.
In total, ten classes for \B backgrounds are defined.

\subsection{Dalitz Plot}
For the decay $B^0 \to K^+ \pi^- \pi^0$, 
with four momentum $p_{B^0}$,
$p_+$, $p_-$ and $p_0$ respectively, the differential decay width with 
respect to the invariant-mass-squared variables
({\em i.e.}, the {\em Dalitz plot variables} (DP)) reads
\begin{equation}
\label{eq:partialWidth}
        d\Gamma(B^0 \rightarrow K^+ \pi^- \pi^0) \;=\;
        \frac{1}{(2\pi)^3}\frac{|\Amptp|^2}{8 m_{B^0}^3}\,ds_{+-} ds_{-0}~,
\end{equation}
where $\Amptp$ is the Lorentz-invariant amplitude
of the three-body decay and 
\begin{equation}
\label{eq:dalitzVariables}
       s_{+-} \;=\; (p_+ + p_-)^2~, \hspace{1cm}
       s_{-0} \;=\; (p_- + p_0)^2~.
\end{equation}

The amplitude $\Amptp$ contains all the underlying dynamics 
of the $\BtoKpipi$ decay. In general, it is the coherent
sum of one nonresonant (NR) term, $\Anonres$, which is assumed 
to be constant in the Dalitz plane, and of several resonant 
amplitudes, $i$, having spin $J$, 
magnitude $a_i$, and phase $\phi_i$:
\begin{equation}
\label{eq:theBigAmplitude}
        \Amptp(s_{+-}\,,s_{-0}) \;=\; a_{\rm NR}e^{i\phi_{\rm NR}}\Anonres 
        \;+
          \sum_i a_i e^{i\phi_i} \cdot \Ares_i(s_{+-}\,,s_{-0})~.
\end{equation}
A similar expression is implied for $\overline{B^0}$ decays, with
$\overline{a}_{\rm NR}$, $\overline{\phi}_{\rm NR}$,
$\overline{a}_i$, and $\overline{\phi}_i$ being assumed different
from $a_{\rm NR}$, $\phi_{\rm NR}$, $a_i$, and $\phi_i$, respectively.

The resonant amplitude $\Ares_i$ is written as a product of four terms
\beqn
\label{eq:fourTerms}
               \Ares_i (\DP) 
        &=&     {^J\!}F_{B,i}           \cdot 
                {^J\!}F_{i}(s)    \cdot 
                {^J\!}K (\DP)      \cdot 
                {^J\!}F_{R,i}(s) ~,
\eeqn
where ${^J\!}F_{B,i}$ is an irrelevant constant form factor for the 
$\Bz$ decay and is absorbed by normalization, 
${^J\!}F_{i}(s)\equiv{^J\!}F(Rq(s))/^J\!F(Rq(m_i^2))$ 
is the ratio of Blatt-Weisskopf damping factors (see below), 
${^J\!}K(\DP)$ is a kinematic function (see below), and $^JF_{R,i}(s)$ 
is a dynamic function describing the resonance (see below).

For a resonance decays to particles $a$ and 
$b$, with invariant mass $\sqrt{s}$, the momentum of the particles $a$ and $b$
in the resonance rest frame, $q$, is given by
\beq
\label{eq:CMmomentum}
        q(s) \;=\; 
                \frac{\sqrt{s}}{2}
                \left(1 - \frac{(m_a + m_b)^2}{s}\right)^{\!\!1/2}        
                \left(1 - \frac{(m_a - m_b)^2}{s}\right)^{\!\!1/2}~,
\eeq
where  $m_a$ and $m_b$ are the masses of particle $a$ and $b$, respectively.

The functions ${^J\!}F(Rq(s))$ are the nuclear 
{\em Blatt-Weisskopf damping factors}~\cite{BlattWeissk}, given by
\beq
        {^0\!}F      \;=\; 1~,\hspace{0.5cm}
        {^1\!}F      \;=\; \frac{1}{\sqrt{1 + R^2 q^2}}~,\hspace{0.5cm}
        {^2\!}F      \;=\; \frac{1}{\sqrt{9 + 3 R^2 q^2 + R^4 q^4}}~,
\eeq
where $R$ is the range parameter. The Blatt-Weisskopf 
damping factors are studied only for systematic 
uncertainty evaluation.

The spin-dependent function is taken to be equal to 1 for a spin-0 resonance,
$-2|{\bf p_d}||{\bf p_b}|\cos{\theta}$ for spin-1, 
and $\frac{4}{3}(|{\bf p_d}||{\bf p_b}|)^2(3\cos^2{\theta} - 1)$ for spin-2,
where ${\bf p_d}$ is the three-momentum of one of the resonance daughters 
and ${\bf p_b}$ is the three-momentum of the bachelor particle, both 
measured in the resonance rest frame, and $\theta$ is the helicity angle 
of the resonance. For a resonance formed from $K^+\pi^0$ ($\pi^-\pi^0$),
the helicity angle is defined by the angle between the $\pi^0$ ($\pi^-$)
in the resonance rest frame and the resonance flight direction in the
$B^0$ rest frame. For a resonance formed from $K^+\pi^-$, the helicity 
angle is defined by the angle between the $K^+$ in the resonance rest 
frame and the resonance flight direction in the $B^0$ rest frame.


Three parameterizations are considered to describe the decay dynamics, 
$^JF_{R,i}(s)$.
Parameters are taken from~\cite{pdg2004} unless 
stated otherwise. The relativistic Breit-Wigner parameterization 
with mass-dependent width is used for
$K^*(892)^{+,0}$, $K_2^*(1430)^{+,0}$, 
and $K^*(1680)^{+,0}$. It is given by
\beq
\label{eq:nominalBW}
        {^J\!}F_{R,i}(s) \;=\; 
                \frac{1}{s - m_i^2 + i m_i{^J}\Gamma_i(s)}~.
\eeq
The $s$-dependent width is defined by
\beq
\label{eq:s-dependentWidth}
        {^J}\Gamma_i(s) \;=\; 
                \Gamma_i^0
                \frac{m_i}{\sqrt{s}}
                \left(\frac{q(s)}{q(m_i^2)}\right)^{\!2J+1}
                \frac{{^J\!}F^2(Rq(s))}{{^J\!}F^2(Rq(m_i^2))}~,
\eeq
where $m_i$ is the mass of the resonance $i$, $\Gamma_i^0=\Gamma_i(m_i^2)$ 
its width.

The Gounaris-Sakurai parameterization~\cite{rhoGS} 
is used to parameterize $\rho(770)^-$, $\rho(1450)^-$ and $\rho(1700)^-$.
For the $K\pi\, S$-wave resonances,
$K_0^*(1430)^{+,0}$, which are found to be dominant in the $K\pi$ invariant
mass range below $2~\gevcc$, an effective-range 
parameterization was suggested~\cite{Estabrooks} to describe the slowly 
increasing phase as a function of $K\pi$ mass.
We use the parameterization as used in the LASS experiment~\cite{LASS},
being modified
for \B decays:
        \beq
          \label{eq:LASS}
           F_{R, i}^{LASS} (s) =  
                      \frac{\sqrt{s}}{ q(s)\,\cot{\delta_B} - i\,q(s)}
                      + 
                      e^{2 i \delta_B}\,\frac{m_i^2\,\Gamma_i^0/q(m_i^2)}
                       {m_i^2 - s - i\, m_i\Gamma_i(s)}\,,
        \eeq

        where
         \begin{equation}
          \label{eq:LASSphase}
           \cot{\delta_B} = \frac{1}{a q(s)} + \frac{1}{2}\,r\,q(s)\,,
         \end{equation}
        $m_i = 1415 \pm 3 \mevcc$,
        $\Gamma_i^0 = 300 \pm 6 \mevcc$, 
        the scattering length $a = 2.07 \pm 0.10 ~(\gevc)^{-1}$, and 
        the effective range $ r = 3.32 \pm 0.34 ~(\gevc)^{-2}$.
 Our results for $K_0^*(1430)$ reported in this 
paper are not purely due to the resonant term, but to the $K\pi$ $S$-wave 
as a whole.

Our nominal model includes the nonresonant contribution and five resonant 
intermediate states: $\rho(770)^-$, $K^*(892)^{+,0}$ and
$K_0^*(1430)^{+,0}$. Variations to this nominal model are used to estimate
the model-dependent systematic uncertainty in the results.

The quasi-two-body fractions and $CP$-violation charge asymmetries are 
extracted as follows:
\begin{eqnarray} \label{eq:PartialFractions}
   f_i\phantom{1} &=& \frac{\int{(\,|a_ie^{i\phi_i}{\cal A}_i|^2 
           + |\overline{a}_ie^{i\overline{\phi}_i} {\cal A}_i|^2}\,)\,{ds_{+-} ds_{-0}}} 
{\int{(\, |\Amptp|^2 + |\Amptpbar|^2 }\,)\,{ds_{+-} ds_{-0}}}~, \\
   A_{\rm CP}^i &=& \frac{|\overline{a_i}|^2 - |a_i|^2}
                         {|\overline{a_i}|^2 + |a_i|^2}~,
\end{eqnarray}
where $a_i$ and $\overline{a_i}$ are the fitted magnitudes, 
$\phi_i$ and $\overline{\phi}_i$ are the fitted relative phases,
for the intermediate state $i$. The extraction is also valid for the 
nonresonant. Due to interference,
the sum of the fractions will in general not add up to unity.

The inclusive branching fraction, ${\cal B}^{incl.}$, and the 
quasi-two-body branching fraction for an intermediate state $i$,
${\cal B}_i$, are given by:
\begin{equation}
   {\cal B}^{incl.} = \frac {N_{sig} } 
                      {{\overline \varepsilon} N_{B{\overline B}}}\,,
   ~~{\cal B}_i = f_i \cdot {\cal B}^{incl.} \,,
\end{equation}
where $N_{sig}$ is the total signal observed in the data,
${\overline \varepsilon}$ is the signal efficiency averaged over the
Dalitz plot, $N_{B{\overline B}}$ is the 
total number of $B{\overline B}$ pairs produced, and $f_i$ is the 
fitted fraction for the intermediate state $i$,
respectively.

\subsection{Maximum Likelihood Fit}
The selected on-resonance data sample is assumed to consist of signal,
continuum-background and $B$-background components. We use the variables
$\mes$, $\deprime$, and the Dalitz plot to discriminate signal from background.
The signal probability density function (PDF) contains 
two parts corresponding respectively to signal events that are 
correctly reconstructed (TM) and signal events that are misreconstructed 
(SCF). A PDF is introduced
to describe the dominant continuum background.
For $B$-related backgrounds, we study the contributions with exclusive 
simulations and group similar contributions into the 
10 $B$ background classes, each of which has an individual PDF in the fit.

The likelihood ${\cal L}$ for $N$ events reads as

\begin{eqnarray}
\label{eq:theLikelihood}
        {\cal L} 
        & = &  e^{-N^{'}} \prod_{i = 1}^{N} \! \bigg\{
                N_{sig}
                \left[  (1-\fscfave){\cal P}_{sig-\TM,i} +
                        \fscfave{\cal P}_{sig-\SCF,i}
                \right]
                \nonumber\\[0.3cm]
        &&
                +\; N_{q\bar q}\frac{1}{2}
                \left(1 + \QKi A_{q\bar q}\right){\cal P}_{q\bar q,i}
                \nonumber \\[0.3cm]
        &&
                +\; \sum_{j=1}^{N^{B^+}_{\rm classes}}
                N_{B^+}^j
                \frac{1}{2}\left(1 + \QKi A_{B^+}^j\right){\cal P}_{B^+,i}^j
                \nonumber \\[0.3cm]
        &&
                +\; \sum_{j=1}^{N^{B^0}_{\rm classes}}
                N_{B^0}^j
                \frac{1}{2}\left(1 + \QKi A_{B^0}^j\right){\cal P}_{B^0,i}^j
                \bigg\}~,
\end{eqnarray}

where,
\begin{itemize}
\item   $N^{'}$ is the sum of all the yields involved, $i.e.$,
        $N_{sig} + N_{q\bar q} + \sum_{j=1}^{N^{B^+}_{\rm classes}}
                N_{B^+}^j + \sum_{j=1}^{N^{B^0}_{\rm classes}}
                N_{B^0}^j$;
\item   $N_{sig}$ is the total number of signal events in the data sample
         to be determined in the fit;

\item   $\fscfave$ is the fraction of the misreconstructed signal events
        averaged over the Dalitz plot,  which is determined
        from the fit;

\item   ${\cal P}_{sig-\TM,i}$ and ${\cal P}_{sig-\SCF,i}$
        are the products of discriminating variable PDFs,
        for the correctly reconstructed and the 
        misreconstructed signal events, respectively;

\item   $N_{q\bar q}$ is the number of the continuum events, which is
        determined in the fit;

\item   $\QKi$ is the kaon charge of the event; we use $q_{\rm K} = 1$ for
        $\Bz$ and $q_{\rm K} = -1$ for $\Bzb$;

\item   $A_{q \bar q}$ parameterizes possible charge
        asymmetry in the continuum events due to detection, reconstruction or
        selection; it is free in the fit;
\item   ${\cal P}_{q\bar q,i}$ is the PDF for the continuum events;

\item   $N^{B^+}_{\rm classes}$ ($N^{B^0}_{\rm classes}$) is the number of
        the charged (neutral) $B$-related background classes considered, 
        equal to 6 (4);

\item   $N_{B^+}^j$ ($N_{B^0}^j$) is the number of events in
        the charged (neutral) $B$-related background class $j$; it is fixed
        to the MC estimate unless stated otherwise;

\item   $A_{B^+}^j$ ( $A_{B^0}^j$ ) describes the charge asymmetry in
        the charged (neutral) $B$ background of class $j$;
        this eventual charge asymmetry could come from physics,
        or detection effects; it is fixed in the fit;

\item   ${\cal P}_{B^+,i}^j$ is the $B^+$-background PDF for class $j$;

\item   ${\cal P}_{B^0,i}^j$ is the $B^0$-background PDF for class $j$;
        two dedicated classes are used for 
        $B^0 \rightarrow {\overline D^0} \pi^0\,,
         {\overline D^0} \rightarrow K^+ \pi^-$ and
        $B^0 \rightarrow D^- K^+\,, D^- \rightarrow \pi^- \pi^0$.
 
\end{itemize}

  Each PDF ${\cal P}_X$ is a product of the PDFs of the four 
discriminating variables, \ie\, 
    $    {\cal P}_{X,i} \;\equiv\;
          P_{X,i}(\mes) \cdot P_{X,i}(\deprime) \cdot P_{X,i}(s_{+-},s_{-0})$.

  The $\mes$ distribution for the correctly reconstructed signal events is 
parameterized using a Crystal Ball function~\cite{CB}, the mean and width 
being simultaneously determined from the fit to data.  
A sum of two Gaussians is used to describe the $\deprime$ distribution, 
the means, widths and fractions being parameterized as linear functions
of the Dalitz plot variable $m^2(K^+\pi^-)$, the intercepts and 
slopes of which are obtained from simulation, and are varied for 
systematic uncertainty
studies. For the misreconstructed signal, the $\mes$ and 
$\deprime$ distributions are both obtained from simulation. 

For the continuum events, the $\mes$ distribution is described by an ARGUS 
shape function~\cite{Argusshape} and its $\deprime$ distribution is 
modeled by a linear function. The shape parameter $\xi$ of the ARGUS 
distribution and the slope of the linear function are determined in a fit
to the on-resonance data but selected with $\mes > 5.23 \gevcc$.
This dedicated fit uses only $\mes$ and $\deprime$ as the discriminating
variables, and considers the signal, the continuum background, and 
the $B$-backgrounds in a similar way as the nominal fit.

The background class $B^0 \rightarrow {\overline D^0} \pi^0 
\,,{\overline D^0} \rightarrow K^+ \pi^-$ shares the parameterizations 
of  $\mes$ and $\deprime$ for the signal events. Its yield is determined 
in the fit to measure the branching fraction. 

The shapes of the other $B$ background $\mes$ - $\deprime$ PDFs are 
obtained from MC simulation and parameterized using two-dimensional empirical 
shape-fitting techniques~\cite{keysPdfs}.

The Dalitz plot PDFs for TM and SCF signal events are

        \beqn
                P_{\TM,i} &=&
                \varepsilon_i\,(1 - \fscfi)\,
                \frac{\AmpAll}{{\cal N}_{\TM}}~,
                \\[0.3cm]
                P_{\SCF,\,i} &=&
                \varepsilon_i\,\fscfi\,
                \frac{\AmpAll}{{\cal N}_{\SCF}}~,
        \eeqn
  where ${\cal N}_{\TM}$ and ${\cal N}_{\SCF}$ are the normalization 
constants to be dynamically determined in the fit, $\varepsilon_i$ and 
$\fscfi$ are the efficiency and the misreconstruction fraction, respectively,
that vary across the Dalitz plot and are obtained with phase-space MC 
simulation. The magnitudes and the relative phases in 
Eq.~\ref{eq:theBigAmplitude} are fit parameters and are allowed 
to vary in the fit, except that we fix $a_i$ to 1.0 and both the phases
to 0 for the $\rho$ component, since this analysis is only sensitive to
relative magnitudes and phases.
The masses and widths for intermediate resonances
are fixed in the fit and are varied by their uncertainties to estimate
the systematic uncertainties in our results.
Because the considered intermediate resonances 
are in general very broad compared to the Dalitz plot resolution, 
we ignore the
resolution effect for the correctly reconstructed 
signal, while we apply a resolution function (matrix) for 
the sizable misreconstructed signal. 

The Dalitz plot PDF for the continuum events is obtained from the
on-resonance sideband events selected with $5.20 < \mes < 5.25 \gevcc$,
plus the off-resonance events selected with $\mes > 5.20 \gevcc$, and
is corrected for the expected $B$-background feed-through in the sideband
of $5.20 < \mes < 5.25 \gevcc$.

For the background class $B^0 \rightarrow {\overline D^0} \pi^0 
\,,{\overline D^0} \rightarrow K^+ \pi^-$, we distinguish the correctly 
reconstructed events and the misreconstructed events. For the correctly 
reconstructed events, a single Gaussian
is used for the reconstructed $D^0$ mass distribution,
the mean and width being free in the fit,
and an empirical $5th$-order polynomial is used to account for 
the reconstruction efficiency effects for the reconstructed $D^0$ helicity 
distribution. 
For the misreconstructed events, a smoothed two-dimensional histogram 
is used to parameterize their contributions in the Dalitz plot, which are 
fixed to the MC expectation. For the other nine classes of $B$-background, 
we model their contributions in the Dalitz plot with
smoothed histograms obtained from simulation.

Multiple solutions occur in the fit for this Dalitz analysis.
To study the effect, we randomize the initial values of the amplitudes and 
phases that are let free in the fit and then the fit is redone many times.
We do observe that there are two sets of solutions that correspond to 
two $\log$-likelihood ($\log{\cal L}$) values. 
The two $\log{\cal L}$ values differ by 0.4 units.
Parameters from the two solutions usually give
very close values except for the phases of $K^*(892)^0$ and $K_0^*(1430)^0$. 
Our strategy is to choose the fit that gives the largest
likelihood to obtain our nominal results. We also report
the other solution for the phases of $K^*(892)^0$ and $K_0^*(1430)^0$.

\section{SYSTEMATIC STUDIES}
\label{sec:Systematics}
The systematic errors in the branching fractions and $CP$-violating 
charge asymmetries, due to assumptions about the decay dynamics, 
are referred to as ``model systematic uncertainty''.
They are obtained by varying the resonance parameters within their 
uncertainties, and by adding less significant resonances to the nominal model.

The other systematic errors in the branching fractions are obtained by adding
in quadrature those evaluated for the signal yield,
the systematic uncertainties in efficiencies of tracking, particle
identification, $\pi^0$ reconstruction and those corresponding to
other selection criteria.
The other systematic errors in the $A_{CP}$ measurements
are introduced by the uncertainties in the treatment of
the $B$ background and by possible charge biases of
the detector.

The systematic uncertainties in the signal yield are primarily due to
the modeling of the signal, of the $B$-background,
and of the continuum in the ML
fit. These are estimated by varying the parameters that are fixed in the 
nominal fit, within their uncertainties. 
The variations in the signal yield are added in quadrature.
The parameters for the signal $\deprime$ PDF are allowed to vary in the 
fit to the data to estimate the corresponding systematic uncertainty.
The expected yields from the background modes are varied according to
the uncertainties in the measured or estimated branching fractions.
Biases observed in toy MC fits are added in quadrature and assigned as 
a systematic uncertainty of the fit procedure, referred to as
``fitting procedure'' in Table~\ref{tab:syssum}.

The basis for evaluating the systematic uncertainties on the cuts that are
applied in the selection process is to study the differences in $\mes$, 
$\deprime$ and the NN output between on-resonance data and Monte Carlo 
simulation.
The corrections and uncertainties in the signal efficiencies are summarized 
in Table~\ref{tab:syssum}.
The parameters for the signal $\mes$ and $\deprime$ PDFs are obtained from 
the fit to the data and are used to estimate the systematic errors due to 
the cuts on $\mes$ and $\deprime$, respectively.
The difference between the data and Monte Carlo distributions
of the NN output is extracted from fully-reconstructed 
$B^+ \rightarrow K^+ \pi^- \pi^0 \pi^+$ decays via intermediate 
$D^0$ or $D^*$, and is used to estimate the systematic error due to the cut on the NN output.

Since \B background modes may exhibit direct \CP violation, the
corresponding parameters (the $A_B^j$'s in Eq.~\ref{eq:theLikelihood}) 
are varied by $\pm0.5$ to conservatively estimate the uncertainties.

Table~\ref{tab:syssum}  summarizes the various sources contributing
to the systematic errors in the branching fractions.
The dominant systematic errors are due to the $\pi^0$ resonstruction,
the $\deprime$ cut, and the cut on the NN output.
Table~\ref{tab:sysacp}  summarizes the possible
sources contributing to the systematic errors in the charge asymmetries.

\begin{table}[pht]
\begin{center}
\caption[Breakdown of systematic errors for the branching fraction measurements.]
        {Breakdown of systematic errors for the branching 
        fraction measurements. The error from each source varies slightly for
        each intermediate state and the most conservative estimate is quoted.}
\label{tab:syssum}
\begin{tabular}{lcc} \hline\hline
&\\[-0.35cm]
\rule[-2.3mm]{0mm}{5mm}     & Charmless yield (events)
& ${\overline D^0} \pi^0$ yield (events) \\ 
&\\[-0.35cm]
\rule[-2.3mm]{0mm}{5mm}
Continuum $\mes$ PDF        & $ 4.3 $  & 0.0 \\
\rule[-2.3mm]{0mm}{5mm}
Continuum $\deprime$ PDF         & $ 2.0 $  & 0.0 \\
\rule[-2.3mm]{0mm}{5mm}
Signal $\deprime$ PDF            & $ 50.0 $ & 12.4 \\
\rule[-2.3mm]{0mm}{5mm}
$B$ backgrounds             & $ 2.7 $ & 11.2 \\
\rule[-2.3mm]{0mm}{5mm}
Fitting procedure           & $ 22.7 $ & 0.0 \\ \hline
&\\[-0.35cm]
\rule[-2.3mm]{0mm}{5mm}
Sub-total (relative)  & $55.2$ ( $4.5\%$ ) & 16.7 ( 3.7\% )\\ \hline 
&\\[-0.35cm]
\rule[-2.3mm]{0mm}{5mm}            
 & \multicolumn{2}{c} {Efficiency and scaling systematics (relative)}\\ 
&\\[-0.35cm]
\rule[-2.3mm]{0mm}{5mm}
Tracking efficiency correction & \multicolumn{2}{c} {$1.6\%$} \\
\rule[-2.3mm]{0mm}{5mm}
PID for tracks                 & \multicolumn{2}{c} {$2.0\%$} \\
\rule[-2.3mm]{0mm}{5mm}
$\pi^0$ reconstruction         & \multicolumn{2}{c} {$6.4\%$} \\
\rule[-2.3mm]{0mm}{5mm}
$\Delta E$ cut efficiency      & \multicolumn{2}{c} {$4.6\%$} \\
\rule[-2.3mm]{0mm}{5mm}
$m_{ES}$   cut efficiency      & \multicolumn{2}{c} {$0.4\%$} \\
\rule[-2.3mm]{0mm}{5mm}
NN  cut efficiency             & \multicolumn{2}{c} {$6.0\%$} \\
\rule[-2.3mm]{0mm}{5mm}
${\rm N}(\BB)$                 & \multicolumn{2}{c} {$1.1\%$} \\ \hline
&\\[-0.35cm]
\rule[-2.3mm]{0mm}{5mm}
Sub-total                      & \multicolumn{2}{c} {$10.3\%$}\\ \hline
&\\[-0.35cm]
\rule[-2.3mm]{0mm}{5mm}
Total systematic error      & $11.2\%$  & $10.9\%$ \\ \hline \hline
\end{tabular}
\end{center}
\end{table}

\begin{table}[pht]
\begin{center}
\caption[Breakdown of systematic errors for the $A_{CP}$ measurements.]
        {Breakdown of systematic errors for the $A_{CP}$ measurements.}
\label{tab:sysacp}
\begin{tabular}{lcccccc} \hline\hline
&\\[-0.35cm]
\rule[-2.3mm]{0mm}{5mm}                                           
                            & $K^*(892)^+$ &  $\rho(770)^-$ 
                            & $K_0^*(1430)^+$ &  $K_0^*(1430)^0$
                            & $K^*(892)^0$ & NR \\\hline
&\\[-0.35cm]
\rule[-2.3mm]{0mm}{5mm}
Detector bias               & \multicolumn{6}{c} {$ 0.01 $}  \\
\rule[-2.3mm]{0mm}{5mm}
$B$ backgrounds             & 0.02 & 0.04 & 0.06 & 0.07 & 0.07 & 0.20 \\\hline
&\\[-0.35cm]
\rule[-2.3mm]{0mm}{5mm}
Total systematic error     & 0.02 & 0.04 & 0.06 & 0.07 & 0.07 & 0.20 \\ \hline \hline
\end{tabular}
\end{center}
\end{table}

\section{RESULTS}
\label{sec:Physics}
The preliminary results are obtained with a fit
to the data sample using our nominal model.
We obtain $1230 \pm 74 $ signal events and 
$454 \pm 24$ $B^0 \rightarrow {\overline D^0} \pi^0, 
{\overline D^0} \rightarrow K^+ \pi^-$ events from the data sample.
The results obtained with the nominal model are shown in 
Table~\ref{tab:results}. The total sum of the fractions obtained with
the nominal fit is $(102.6 \pm 8.0)\%$. Table~\ref{tab:resultsMinor} 
shows the results for the less significant modes, obtained by adding in turn 
one mode to the nominal fit. The results obtained from these altered fits, 
for the modes in the nominal model, are compatible.
 
The inclusive signal efficiency is estimated to be 
$(16.6 \pm 0.1)\%$ 
with the observed Dalitz plot structure. 
We measure the inclusive branching fraction to be:

\begin{equation}
   {\cal B} (B^0 \rightarrow K^+ \pi^- \pi^0) \,=\, 
      ( 34.9 \pm 2.1 (stat.) \pm 3.9 (syst.)) \times 10^{-6}\,.
\end{equation}

We find 
\begin{equation}
   {\cal B} (B^0 \rightarrow {\overline D^0} \pi^0) \,=\, 
      ( 3.3 \pm 0.2 (stat.) \pm 0.4 (syst.)) \times 10^{-4}\,.
\end{equation}

The signal significance for ${\cal B}(B^0 \rightarrow K^*(892)^0 \pi^0)$, 
including statistical and systematic errors, is $4.2 \sigma$.
We set an upper limit of 
${\cal B}(B^0 \rightarrow K^+ \pi^- \pi^0~nonresonant) < 4.7\times10^{-6}$
at $90\%$ confidence level, with systematic errors being 
taken into account.

There exists another solution with a $\log{\cal L}$ 0.4 units lower than
the maximum $\log{\cal L}$. This solution
gives very compatible results except for
the relative phases of $K_0^*(1430)^0$ and $K^*(892)^0$.

\begin{figure}[!htb]
\begin{center}
\includegraphics[height=8cm]{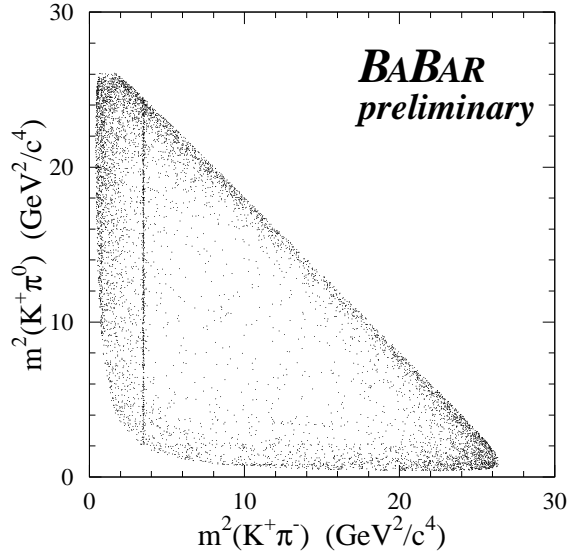}
\caption{Dalitz plot for the selected $K^+ \pi^- \pi^0$  events. 
         The visible narrow band comes from the 
         $B^0 \rightarrow \overline{D}^0 \pi^0$ decays.}
\label{fig:DalitzPlot}
\end{center}
\end{figure}

\begin{figure}[!htb]
\begin{center}
\includegraphics[height=7.5cm]{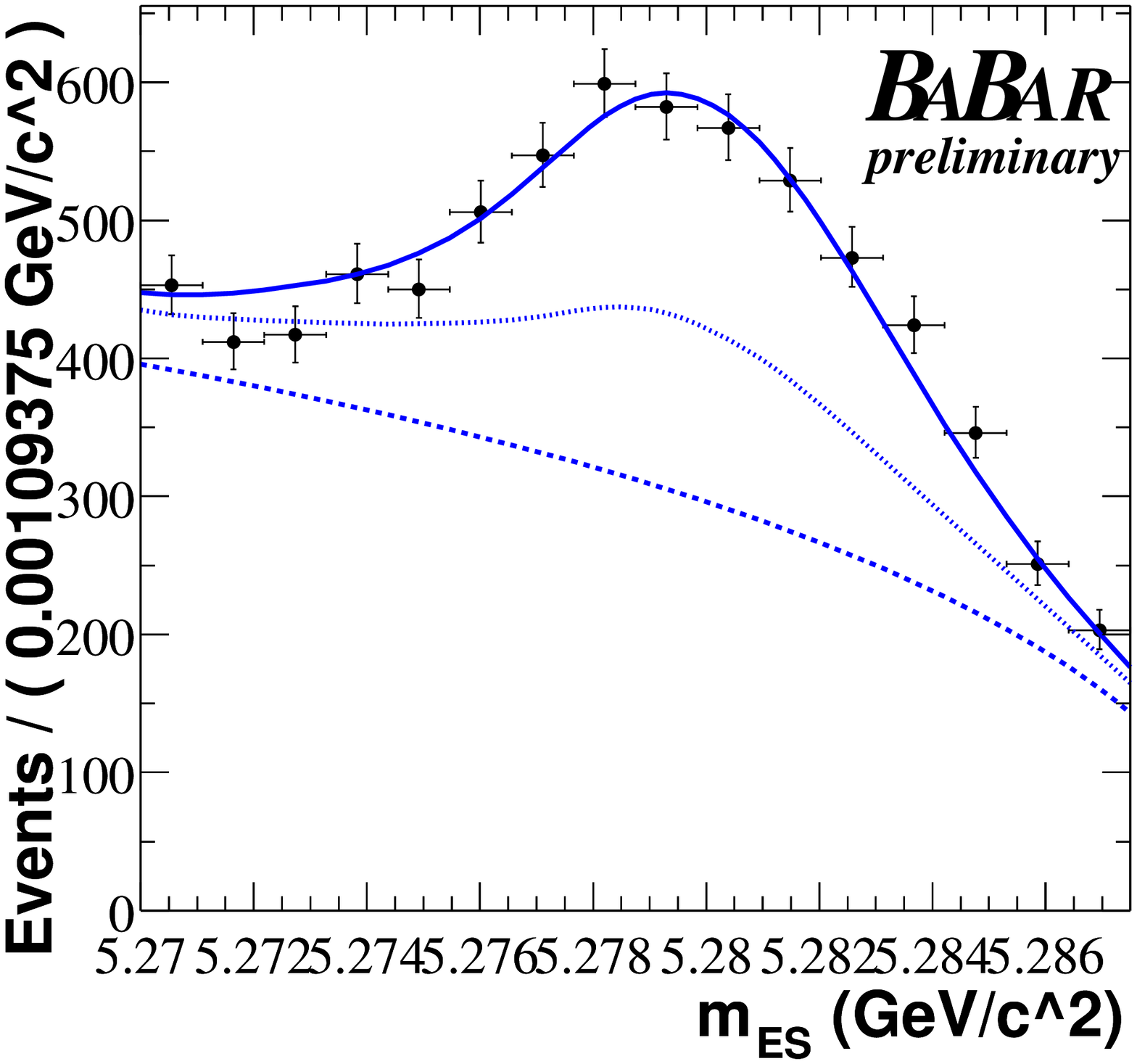}
\includegraphics[height=7.5cm]{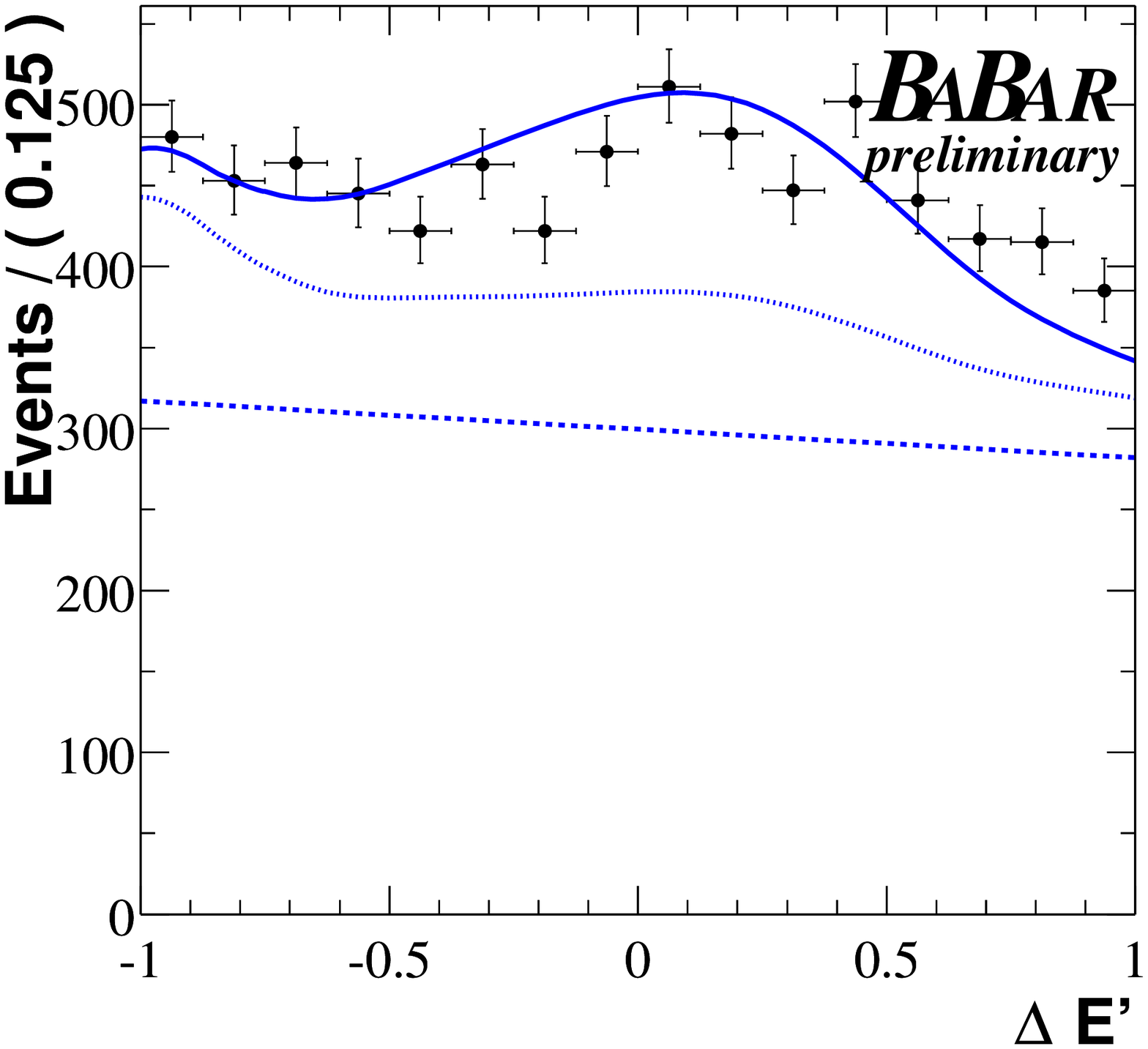}
\caption{$\mes$ (left) and $\deprime$ (right) projections for the 
         selected $K^+ \pi^- \pi^0$  events. 
         The data are the points with error bars. The fit result is the 
         (top) solid line. The (bottom) dashed line represents the 
         continuum background and the (middle) dotted line is the 
         \B\-background added on top of the continuum background.
         Note that the \B\-backgound includes the
         $B^0 \rightarrow \overline{D}^0 \pi^0$, 
         $\overline{D}^0 \to K^+ \pi^-$decays.}
\label{fig:mesde}
\end{center}
\end{figure}

\begin{figure}[!htbp]
\begin{center}
\includegraphics[height=7.0cm]{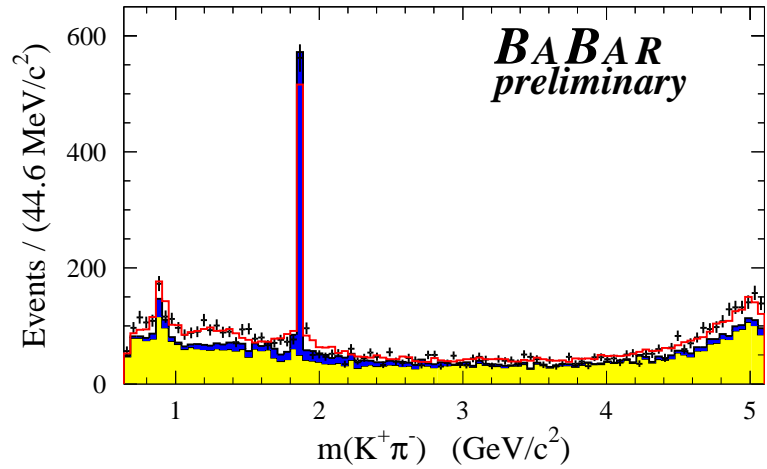}
\includegraphics[height=7.0cm]{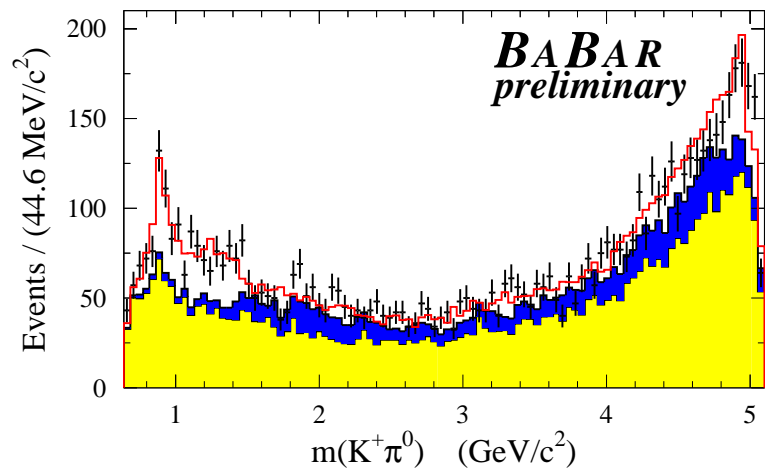}
\includegraphics[height=7.0cm]{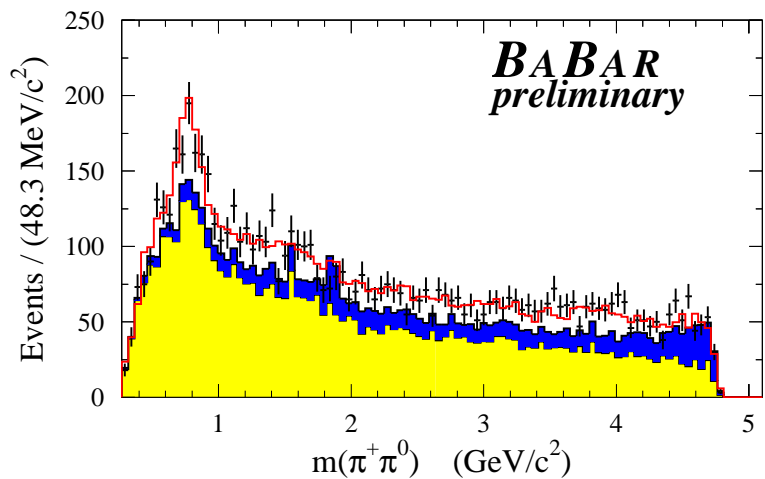}
\vspace{-0.8cm}
\caption{Projection plots for the mass pairs. The data are the black points
with error bars (statistical only).
The total fit result is the solid (red) line. The continuum background is 
the light shaded (yellow) and the $B$-background is the dark shaded (blue) 
added on top of the continuum background.}
\label{fig:Projections}
\end{center}
\end{figure}

The Dalitz plot for the selected $K^+ \pi^- \pi^0$ events is shown in 
Fig.~\ref{fig:DalitzPlot}. 
The $\mes$ and $\deprime$ projections are shown in Fig.~\ref{fig:mesde}.
Projection plots of the invariant mass pairs are shown in 
Fig.~\ref{fig:Projections}. 

\begin{table}[!htb]
\caption{Results obtained with a fit to the data using the nominal model. 
         The first errors in the table are statistical while the second 
         and the third (when present) are 
         systematic errors as discussed in Sec.~\ref{sec:Systematics}.
         The last column indicates the likelihood
         change when the mode is omitted in turn in the nominal model.
         $f$ and $A_{CP}$ have been defined in Eq.~\ref{eq:PartialFractions}, 
         while $\phi$ and ${\overline \phi}$ have been defined in 
         Eq.~\ref{eq:theBigAmplitude}. For the other solution, 
         $\phi_{K_0^*(1430)^0} = (106 \pm 44)^\circ $,
         $\phi_{K^*(892)^0} = (-44 \pm 48)^\circ $, and all the other parameters
         give very compatible results. }
\vspace{-0.1cm}
\begin{center}
\begin{tabular}{l|c|c|c|c} \hline
&&&\\[-0.35cm]
\rule[-2.3mm]{0mm}{5mm}                                           
Mode & $f$ (\%) & $\phi$ / ${\overline \phi}$ (degree) 
& $A_{CP}$ & $\Delta \log{\cal L}$
 \\\hline
&&&\\[-0.35cm]
\rule[-2.3mm]{0mm}{5mm}                                           
$ K^*(892)^+ \pi^-$ 
& $10.4_{-2.0}^{+2.1} \pm 0.8 $ & $138 \pm 35$ /  
& $-0.25 \pm 0.17 \pm 0.02 \pm 0.02 $ & -42.5\\
& & $174 \pm 42$ & &\\\hline
&&&\\[-0.35cm]
\rule[-2.3mm]{0mm}{5mm}                                           
$ \rho(770)^- K^+ $ 
& $24.6_{-2.9}^{+3.6} \pm 0.6$  & $0.0$ (fixed) /  
 & $0.13^{+0.14}_{-0.17} \pm 0.04 \pm 0.13 $ & -48.3\\
& & $0.0$ (fixed) && \\\hline
&&&\\[-0.35cm]
\rule[-2.3mm]{0mm}{5mm}                                           
$ K_0^*(1430)^+ \pi^-$ 
& $32.2\pm 3.8 \pm 9.4$ & $115 \pm 34$ /
& $-0.07 \pm 0.12 \pm 0.06 \pm 0.05 $ & -92.9  \\
& &  $149 \pm 35$ & & \\\hline
&&&\\[-0.35cm]
\rule[-2.3mm]{0mm}{5mm}                                           
$ K_0^*(1430)^0 \pi^0$ 
& $22.5 \pm 4.0 \pm 7.2$ & $-12 \pm 40$ / 
& $-0.34 \pm 0.15 \pm 0.07 \pm 0.08 $ & -71.4 \\
& & $8 \pm 42$ & & \\\hline
&&&\\[-0.35cm]
\rule[-2.3mm]{0mm}{5mm}                                           
$ K^*(892)^0 \pi^0$ 
& $5.8_{-1.5}^{+1.7} \pm 0.6$ & $-160 \pm 41$ /  
& $-0.01^{+0.24}_{-0.22} \pm 0.07 \pm 0.11$  & -23.8 \\ 
& & $-144 \pm 41$ & & \\\hline
&&&\\[-0.35cm]
\rule[-2.3mm]{0mm}{5mm}                                           
$ Nonresonant$ 
& $7.1_{-2.9}^{+3.6} \pm 0.1$ & $55 \pm 28$ / 
& $-0.12^{+0.36}_{-0.37} \pm 0.20 \pm 0.07$ & -7.3\\
& & $79 \pm 28$ & & \\\hline
\end{tabular}
\end{center}
\label{tab:results}
\end{table}

\begin{table}[!phtb]
\caption{Results obtained for the less significant intermediate states 
         when they are added to the nominal model in turn. 
         The last column indicates the likelihood change.}
\begin{center}
\begin{tabular}{l|c|c} \hline
&&\\[-0.35cm]
\rule[-2.3mm]{0mm}{5mm}                                           
Mode & Fraction (\%) 
& $\Delta \log{\cal L}$
 \\\hline
&&\\[-0.35cm]
\rule[-2.3mm]{0mm}{5mm}                                           
$ K_2^*(1430)^+ \pi^- $ 
& $3.5_{-1.5}^{+2.0} $ 
& 7.3 \\\hline
&&\\[-0.35cm]
\rule[-2.3mm]{0mm}{5mm}                                           
$ K_2^*(1430)^0 \pi^0 $ 
& $1.6_{-1.4}^{+1.7}$ 
& 3.4\\\hline
&&\\[-0.35cm]
\rule[-2.3mm]{0mm}{5mm}                                           
$ K^*(1680)^+ \pi^- $ 
& $3.9_{-1.7}^{+2.3}$ 
& 5.8 \\\hline
&&\\[-0.35cm]
\rule[-2.3mm]{0mm}{5mm}                                           
$ K^*(1680)^0 \pi^0 $ 
& $0.8_{-0.8}^{+1.4} $ 
& 1.6\\\hline
&&\\[-0.35cm]
\rule[-2.3mm]{0mm}{5mm}                                           
$ \rho(1450)^- K^+ $ 
& $5.0_{-2.7}^{+2.8}$ 
& 3.6 \\\hline
&&\\[-0.35cm]
\rule[-2.3mm]{0mm}{5mm}                                           
$ \rho(1700)^- K^+ $ 
& $2.5_{-1.5}^{+1.8}$ 
& 2.1\\\hline
\end{tabular}
\end{center}
\label{tab:resultsMinor}
\end{table}

\section{SUMMARY}
\label{sec:Summary}
Due to lack of knowledge of the final state interactions in 
$B \rightarrow K^+ \pi^- \pi^0$ decays, this Dalitz analysis assumes 
uniform phase space for the nonresonant decay amplitude and utilizes 
parameterizations obtained from non-$B$-meson experiments to 
model the intermediate resonant amplitudes.
Non-unique parameterizations of the decay amplitude are covered by
the systematic uncertainties.

\vspace{0.2cm}
A comparison with other measurements is summarized in 
Table~\ref{tab:comparison}.

\begin{table}[p]
\begin{center}
\caption[Comparison]
 {Comparison between measurements of branching fractions 
  ($\times 10^{-6}$)
  and $CP$-violating charge asymmetries. Note that Belle measured  
  $K_X^*$, which is defined to have $1.1\gevcc < m(K^+\pi^{-,0}) < 1.6\gevcc$
  and has a helicity angular distribution indicating
  a scalar behavior. Our preliminary results in this table are 
  corrected for the
  secondary decay branching fractions except for those indicated by a 
  {\bf $\dagger$}. 
  Previously published results from \babar\
  on $\rho^- K^+$ (listed within parentheses) were 
  obtained with a subset of $81~fb^{-1}$ of the data 
  used for this analysis.}
\label{tab:comparison}
\begin{tabular}{l|c|c|c}
\hline
\hline
  & CLEO & Belle & \babar \\\hline
&&&\\[-0.35cm]
\rule[-2.3mm]{0mm}{5mm}                                           
${\cal B} (B^0 \rightarrow K^+ \pi^- \pi^0 ~{\rm inclusive})$& $<40$ 
     & $36.6 \pm 4.2 \pm 3.0$ & $34.9 \pm 2.1 \pm 3.9$ \\
$A_{\rm CP} (B^0 \rightarrow K^+ \pi^- \pi^0 ~{\rm inclusive})$& - 
     & $0.07 \pm 0.11 \pm 0.01$ & - \\\hline
&&&\\[-0.35cm]
\rule[-2.3mm]{0mm}{5mm}                                           
${\cal B} (B^0 \rightarrow K^*(892)^+\pi^-)$ & $16\pm6\pm2$ 
     & $14.8 \pm 4.6 \pm 1.5 \pm 2.4$ & $10.9 \pm 2.3 \pm 1.5$\\
$A_{\rm CP} (B^0 \rightarrow K^*(892)^+\pi^-)$ 
     & $0.26 ^{+0.33+0.19}_{-0.34-0.08}$ 
     &- & $-0.25 \pm 0.17 \pm 0.02 \pm 0.02$ \\\hline
&&&\\[-0.35cm]
\rule[-2.3mm]{0mm}{5mm}                                           
${\cal B} (B^0 \rightarrow K^*(892)^0\pi^0) $& $<3.6$ 
     & $<3.5$ & $3.0 \pm 0.9 \pm 0.5$\\
$A_{\rm CP} (B^0 \rightarrow K^*(892)^0\pi^0) $& -
     & - & $-0.01 _{-0.22}^{+0.24} \pm 0.07 \pm 0.11$ \\\hline
&&&\\[-0.35cm]
\rule[-2.3mm]{0mm}{5mm}                                           
${\cal B} (B^0 \rightarrow \rho(770)^- K^+)$& $16 \pm 8 \pm3$
     & $15.1 \pm 3.4 \pm 1.5 \pm 2.1$
     & $8.6 \pm 1.4 \pm 1.0$ \\
     & 
     & 
     & ($7.3 \pm 1.3 \pm 1.3$) \\
$A_{\rm CP} (B^0 \rightarrow \rho(770)^- K^+)$& -
     & $0.22 ^{+0.22+0.06}_{-0.23-0.02}$
     & $0.13 ^{+0.14}_{-0.17} \pm 0.04 \pm 0.13$ \\
     & 
     & 
     & ($0.18 \pm 0.12 \pm 0.02$) \\\hline
&&&\\[-0.35cm]
\rule[-2.3mm]{0mm}{5mm}                                           
${\cal B} (B^0 \rightarrow K_0^*(1430)^{0}\pi^0 )^\dagger$
& - & $6.1^{+1.6+0.5}_{-1.5-0.6}$ & $7.9 \pm 1.5 \pm 2.7$\\
$A_{\rm CP} (B^0 \rightarrow K_0^*(1430)^{0}\pi^0 )$
&-& - & $-0.34\pm0.15\pm0.07\pm0.08$ \\\hline
&&&\\[-0.35cm]
\rule[-2.3mm]{0mm}{5mm}                                           
${\cal B} (B^0 \rightarrow K_0^*(1430)^+\pi^- )^\dagger$ 
&-& $5.1\pm1.5_{-0.7}^{+0.6}$ & $11.2 \pm 1.5 \pm 3.5$ \\
$A_{\rm CP} (B^0 \rightarrow K_0^*(1430)^+\pi^- )$ 
&- & - &  $-0.07 \pm 0.12 \pm 0.06 \pm 0.05$\\\hline
&&&\\[-0.35cm]
\rule[-2.3mm]{0mm}{5mm}                                           
${\cal B} (B^0 \rightarrow K^+ \pi^- \pi^0 ~{\rm NR})$ & -
     & $< 9.4$
     &  $ < 4.6$\\\hline

&&&\\[-0.35cm]
\rule[-2.3mm]{0mm}{5mm}                                           
${\cal B} ( B^0 \rightarrow K_2^*(1430)^+ \pi^- )^\dagger$ 
& - & - & $ <2.2 $ \\\hline
&&&\\[-0.35cm]
\rule[-2.3mm]{0mm}{5mm}                                           
${\cal B} ( B^0 \rightarrow K_2^*(1430)^0 \pi^0 )^\dagger$ 
& - & - & $ < 1.2$ \\\hline
&&&\\[-0.35cm]
\rule[-2.3mm]{0mm}{5mm}                                           
${\cal B} ( B^0 \rightarrow K^*(1680)^+ \pi^- )^\dagger$ 
& - & -  & $ < 2.5$ \\\hline
&&&\\[-0.35cm]
\rule[-2.3mm]{0mm}{5mm}                                           
${\cal B} (B^0 \rightarrow  K^*(1680)^0 \pi^0 )^\dagger$ 
& - & -  & $ < 1.3$ \\\hline
&&&\\[-0.35cm]
\rule[-2.3mm]{0mm}{5mm}                                           
$ {\cal B} (B^0 \rightarrow \rho(1450)^- K^+ )^\dagger$ 
& - & - & $ < 3.2$ \\\hline
&&&\\[-0.35cm]
\rule[-2.3mm]{0mm}{5mm}                                           
${\cal B} (B^0 \rightarrow  \rho(1700)^- K^+ )^\dagger$ 
& - & - & $ < 1.7$ \\\hline

\end{tabular}
\end{center}
\end{table}

The measured branching fraction of $B^0 \rightarrow {\overline D^0} \pi^0$
is in good agreement with the current world average,
$(2.7 \pm 0.8)\times 10^{-4}$~\cite{pdg2004}.

The measured branching fractions and charge asymmetries of the decays 
$B^0 \rightarrow \rho(770)^- K^+$ and $B^0 \rightarrow K^*(892)^+ \pi^-$ 
are in agreement with the previously reported 
measurements~\cite{CLEOkpipi0, BaBarRhopi, BELLEkpipi0}. 

We report observations of the intermediate $S$-wave decays,
$B^0 \rightarrow K_0^*(1430)^+ \pi^-$ and 
$B^0 \rightarrow K_0^*(1430)^0 \pi^0$, with the 
LASS parameterization~\cite{LASS}.
We do not see any significant direct $CP$ violation in these two decay 
modes.

We observe the first evidence of $B^0 \rightarrow K^*(892)^0 \pi^0$ decay 
with a branching fraction in agreement with the reported upper 
limit~\cite{BELLEkpipi0}. We do not see any direct 
$CP$ violation in this decay mode.

We set upper limits on the branching 
fractions of the nonresonant decay and the other six less significant 
intermediate states.

\section{ACKNOWLEDGMENTS}
\label{sec:Acknowledgments}

\input pubboard/acknowledgements

\end{document}

%% file: pubboard/authors_sum2004.tex
\begin{center}
\small

The \babar\ Collaboration,
\bigskip

%
B.~Aubert,
R.~Barate,
D.~Boutigny,
F.~Couderc,
J.-M.~Gaillard,
A.~Hicheur,
Y.~Karyotakis,
J.~P.~Lees,
V.~Tisserand,
A.~Zghiche
\inst{Laboratoire de Physique des Particules, F-74941 Annecy-le-Vieux, France }
% ...

\end{center}\newpage

%% file: pubboard/acknowledgements.tex
We are grateful for the 
extraordinary contributions of our \pep2\ colleagues in
achieving the excellent luminosity and machine conditions
that have made this work possible.
The success of this project also relies critically on the 
expertise and dedication of the computing organizations that 
support \babar.
The collaborating institutions wish to thank 
SLAC for its support and the kind hospitality extended to them. 
This work is supported by the
US Department of Energy
and National Science Foundation, the
Natural Sciences and Engineering Research Council (Canada),
Institute of High Energy Physics (China), the
Commissariat \`a l'Energie Atomique and
Institut National de Physique Nucl\'eaire et de Physique des Particules
(France), the
Bundesministerium f\"ur Bildung und Forschung and
Deutsche Forschungsgemeinschaft
(Germany), the
Istituto Nazionale di Fisica Nucleare (Italy),
the Foundation for Fundamental Research on Matter (The Netherlands),
the Research Council of Norway, the
Ministry of Science and Technology of the Russian Federation, and the
Particle Physics and Astronomy Research Council (United Kingdom). 
Individuals have received support from 
CONACyT (Mexico),
the A. P. Sloan Foundation, 
the Research Corporation,
and the Alexander von Humboldt Foundation.

%% file: paper.bbl
\begin{thebibliography}{99}

\bibitem{CKM}           N.~Cabibbo, \jprl{10}, 531 (1963); 
                        M.~Kobayashi, T.~Maskawa, 
                         \progtp{49}, 652 (1973).

\bibitem{Gronau}        C.W.\ Chiang, M.\ Gronau, Z.L.\ Luo, J.L.\ Rosner, 
                        D.A\ Suprun, Phys. Rev. {\bf D69}, 034001(2004).

\bibitem{Deshpande_PEW} N.G.\ Deshpande and X.\ He, Phys. Rev. Lett. 
                        {\bf 74}, 26 (1995); \\
			R.~Fleisher, Z. Phys. {\bf C62}, 81 (1994).
  
\bibitem{CLEOkpipi0}    CLEO Collaboration, E.~Eckhart \ea,
                        Phys. Rev. Lett. {\bf 89}, 251801 (2002).

\bibitem{BaBarRhopi}   \babar\ Collaboration, B.\ Aubert \ea,
                       \jprl{91}, 201802 (2003).
\bibitem{BELLEkpipi0}   Belle Collaboration, P.~Chang \ea, 
                        BELLE-CONF-0317, hep-ex/0406075 (2003).

\bibitem{ref:babar}     \babar\ Collaboration, B.\ Aubert {\em et al.},
                        Nucl.\ Instrum.\ Methods {\bf A479}, 1-116 (2002).

\bibitem{BlattWeissk}   J.~Blatt and V.~Weisskopf, {\em ``Theoretical 
                        Nuclear Physics''}, John Wiley \& Sons, New York,
                        1956.

\bibitem{pdg2004}       Particle Data Group, S.~Eidelman {\em et al.}, 
                        Phys.\ Lett.\ {\bf B592}, 1 (2004).

\bibitem{rhoGS}         G.J.~Gounaris and J.J.~Sakurai,
                        Phys. Rev. Lett. {\bf 21}, 244 (1968).


\bibitem{Estabrooks}    P.\ Estabrooks, Phys. Rev. {\bf D19}, 2678 (1979).

\bibitem{LASS}          D.\ Aston \ea, Nucl. Phys. {\bf B296}, 493 (1988).

\bibitem{CB}            J.E.~Gaiser {\em et al.}, 
                        Phys. Rev. {\bf D34}, 711 (1986).

\bibitem{Argusshape}    ARGUS Collaboration, H.~Albrecht {\em  et al.}, 
                        Z. Phys. {\bf C48}, 543 (1990).

\bibitem{keysPdfs}      K.S.~Cranmer, Comput. Phys. Commun. {\bf 136}, 
                        198 (2001), hep-ex/0011057, ALEPH 99-144 (1999).

\end{thebibliography}
